# Compressed Sensing for Denoising in Adaptive System Identification

Seyed Hossein Hosseini and Mahrokh G. Shayesteh
Department of Electrical Engineering, Urmia University, Urmia, Iran
Emails: st_h.hosseini@urmia.ac.ir ; m.shayesteh@urmia.ac.ir

*Abstract*— We propose a new technique for adaptive identification of sparse systems based on the compressed sensing (CS) theory. We manipulate the transmitted pilot (input signal) and the received signal such that the weights of adaptive filter approach the compressed version of the sparse system instead of the original system. To this end, we use random filter structure at the transmitter to form the measurement matrix according to the CS framework. The original sparse system can be reconstructed by the conventional recovery algorithms. As a result, the denoising property of CS can be deployed in the proposed method at the recovery stage. The experiments indicate significant performance improvement of proposed method compared to the conventional LMS method which directly identifies the sparse system. Furthermore, at low levels of sparsity, our method outperforms a specialized identification algorithm that promotes sparsity.

## I. INTRODUCTION

A basic implementation for adaptive system identification is achieved by the least mean square (LMS) algorithm [1]. Although this algorithm is able to identify any unknown systems, but a priori knowledge about the desired system can be helpful for performance improvement. A common priori information is sparsity. A system that its taps are modeled by the vector $h=[h_1,h_2,...,h_N]^T$ is $k$-sparse, if the number of its non-zero taps is at most $k$, such that $k \ll N$. Some adaptive algorithms have been developed in the literature to exploit sparsity. The strategy of these algorithms can be classified into two main categories: *a)* manipulation of the step size [2] and *b)* regularization of error function by adding another function that promotes the sparsity of solution [3], [4]. The second category is affected by emerging compressed sensing (CS) theory.

Initially, CS theory was proposed for sampling of sparse signals. However, nowadays it is applied in various applications. In signal acquisition framework, CS measures with the sub-Nyquist rate and reconstructs the original Nyquist samples by the optimization tools. In mathematical expression, if the vector $s=[s_1,s_2,...,s_N]^T$ is the Nyquist samples of sparse signal $s$, the vector $y=[y_1,y_2,...,y_M]^T$ carries the sufficient information about $s$ ($M<N$) which is obtained by the linear measurements as [5]:

$$y = \Phi s \quad (1)$$

where $\Phi \in R^{M \times N}$ is the measurement matrix. In the reconstruction stage (recovery of $s$ from known $y$ and $\Phi$), CS promotes sparsity of the solution by adding a regularization function ($l_1$ norm) to the least square error:

$$\hat{s} = \arg\min_s \frac{1}{2}\|y-\Phi s\|_2^2 + \lambda\|s\|_1 \quad (2)$$

where $\|.\|_1$ indicates the $l_1$ norm of the signal which is defined as $\|s\|_1 = |s_1|+|s_2|+\cdots+|s_N|$, $\|.\|_2$ denotes the $l_2$ norm or amplitude of the vector, and $\lambda$ is the regularization parameter. The authors in [3] and [4] use $l_1$ regularizer and similar types to encourage sparse solutions in the adaptive identification by LMS and RLS algorithms, respectively.

Although $l_1$ minimization always leads to the sparse solution but it does not guarantee the exact solution of the undetermined system mentioned in (1). CS theory provides a performance guarantee by exerting the restricted isometry property (RIP) constraint on the measurement matrix $\Phi$ [6], [7]. CS also shows that some random matrices such as Gaussian and Bernoulli satisfy RIP for any sparse vector with the overwhelming probability, if the number of measurements ($M$) is of order of the information of sparse signal, i.e.

$$M \geq O(k \log N) \quad (3)$$

Similarly, RIP is a sufficient condition to guarantee the recovery by two other classes of the sparse reconstruction algorithms: greedy [8] and Bayesian [9] methods.

In this paper, we propose a new method, in which the adaptive filter identifies the compressed version of the sparse system $h$. For this purpose, we deal with the time variant feature of compressed sensing measurements. The identification in the reduced dimension (compressive identification) has several advantages over the other methods. As an important advantage, utilization of an appropriate CS algorithm at the reconstruction stage of final identification, results in a significant denoising performance which leads to an improvement in the distortion estimation compared to the conventional direct method LMS and the mentioned specialized algorithms. Moreover, in the proposed method, exploiting extra information about system will be easily possible by the state-of-the-art CS algorithms.

The rest of this paper is organized as follows. In Section II, we briefly review the conventional model of adaptive identification, the sparse LMS algorithm, and a specific kind of compressive measurement. In Section III, we explain the proposed method for identification in the reduced dimension, i.e. compressive identification. In Section IV, we explore the advantages of this type of identification by performance comparison (before and after reconstruction) with the conventional and specialized methods. Finally, conclusion is presented in Section V.

## II. SYSTEM MODEL

In this section, we review the notations, algorithms, and measurement used in the paper.

### A. Identification Algorithms

*1) Conventional method (LMS):* Fig. 1 shows the conventional structure of direct identification in the time domain. The taps of the unknown system are modeled by an *N*-element vector *h*. The pilot (input) signal used in the *n*-th iteration of identification is shown by $x^{(n)} = [x_n, x_{n-1}, ..., x_{n-N+1}]^T$. Further, the weights of adaptive filter are denoted by vector $w^{(n)} = [w_1^{(n)}, w_2^{(n)}, ..., w_N^{(n)}]^T$ which has the same length *N* as the impulse response of the system *h*. Hence, the output of filter in the *n*-th iteration will be as follows:

$$u^{(n)} = (w^{(n)})^T x^{(n)} \quad (4)$$

The received signal that is considered as the desired signal is:

$$d^{(n)} = h^T x^{(n)} + v^{(n)} \quad (5)$$

where $v^{(n)}$ is the additive white Gaussian noise (AWGN).

We utilize the least mean square (LMS) error as the cost function to derive an update rule for the weights, i.e.:

$$F(n) = \frac{1}{2}(e^{(n)})^2 \quad (6)$$

where $e^{(n)} = u^{(n)} - d^{(n)}$. Hence, the weights are updated as follows:

$$w^{(n+1)} = w^{(n)} - \mu e^{(n)} x^{(n)} \quad (7)$$

where $\mu$ is the step size.

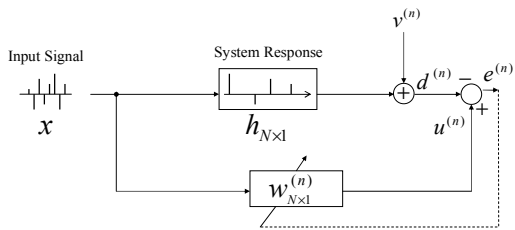

Figure 1. Conventional scheme for system identification.

*2) ZA-LMS:* To compare the proposed method with the dedicated algorithms, we will use following update which promotes the sparsity of the solution by Zero-Attracting property of the *sign* function (ZA-LMS algorithm) [3]:

$$w^{(n+1)} = w^{(n)} - \mu e^{(n)} x^{(n)} - \rho \, sign(w^{(n)}) \quad (8)$$

where $\rho$ is the regularization parameter and *sign* is derivation of $l_1$ regularizer defined as follows

$$sign(x) = \begin{cases} x/|x| & x \neq 0 \\ 0 & x = 0 \end{cases} \quad (9)$$

### B. Measurement Matrix

Random filter is one of the structured compressive measurements which has more feasible implementation than the initially proposed random matrices. A finite impulse response filter (FIR) with length *L*, i.e. $f = [f_1, f_2, ..., f_L]$, is considered as a random filter, if its taps have random values [10]. Fig. 2 shows the structure of signal acquisition using a random filter. When the input of the random filter is the samples of the sparse signal $s[n]$, then its decimated output $s_f[qn]$ is the compressive information of the input such that:

$$s_f[n] = s[n] * f[n] \quad (10)$$

where *q* is the decimation rate [11].

In this structure, the measurement matrix $\Phi_f$ is the convolution matrix corresponding to the random filter which its rows are eliminated at the rate of decimation. Hence, the number of measurements is equal to $\lfloor (N+L-1)/q \rfloor$.

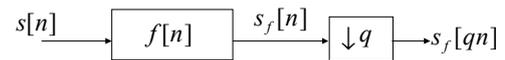

Figure 2. Signal acquisition by random filter.

## III. PROPOSED SCHEME OF IDENTIFICATION

In the proposed method, we identify the sparse system in the compressed domain, i.e. the weights of filter approximate the $M \times 1$ vector $\Phi h$ instead of the $N \times 1$ vector *h* in the conventional method ($M < N$). The original system is recovered by a sparse reconstruction algorithm that has acceptable denoising performance.

Compressive identification requires some changes at the transmitter-receiver front ends, such that the received signal (desired signal) should be as:

$$d^{(n)} = (\Phi h)^T x^{(n)} + v^{(n)} \quad (11)$$

Further, any modification must preserve the linear time invariant (LTI) property of the system. Although the measurement stage in the compressed sensing is linear, but there is not any time invariant structure in this framework. Nevertheless, the construction of such LTI system is possible by using additional operators.

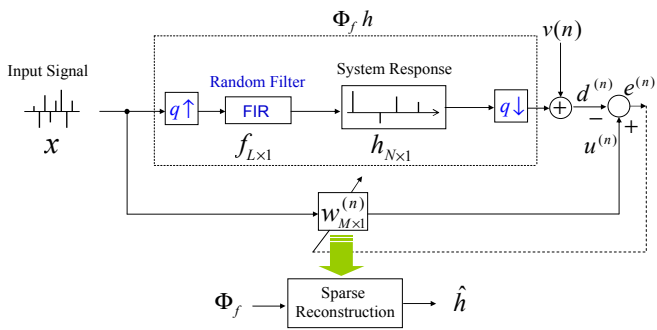

Figure 3. The proposed process for compressive sparse system identification.

The proposed structure for compressive identification is modeled as shown in Fig. 3. We use the random filter structure to form the compressed version of $h$ ( $\Phi_f h$ ). The corresponding FIR filter is added to the transmitter front end and the decimator is applied at the receiver. The overall system would not be time invariant because of the decimator at the receiver. In order to solve this problem, it is sufficient to interpolate the transmitted pilot (input) signal at the transmitter with the same rate of decimator. Consequently, the overall system would be time invariant.

## IV. EXPERIMENTS AND RESULTS

In this section, we present two experiments to investigate the advantages of the proposed method. In the first experiment, we compare the performance of the identification in the reduced dimension (compressive identification) with the conventional LMS method. In the second experiment, we validate the main advantage of the proposed method by denoising capacity of a compressed sensing algorithm and compare the final performance with LMS and ZA-LMS methods.

### A. Compressive Identification

Here, we compare the distortion and the number of iterations in the two cases of conventional and compressive identifications (without reconstruction). The criterion for performance evaluation in the conventional method is the mean of distortion as $E(\|h-\hat{h}\|_2^2 / \|h\|_2^2)$ and in the compressive method, it is $E(\|\Phi_f h - \widehat{\Phi_f h}\|_2^2 / \|\Phi_f h\|_2^2)$. The operator $E(.)$ denotes the expectation of distortion which was achieved by averaging over 100 trials. In each trial, a random impulse response was generated in which the positions of the $k$ non-zero elements were chosen uniformly and their values were selected from a Gaussian distribution with zero mean and unit variance. Further, the amplitudes of the taps of the random filter, and the samples of pilot signal $x$ were selected from the same i.i.d. Gaussian distribution.

The lengths of the impulse response $h$ and random filter $f$ were set to $N=500$ and $L=80$, respectively. The number of non-zero elements of the system $h$ was set to $k=40$. The pilot data were interpolated and decimated at the transmitter and receiver with the rate $q=2$. Hence, the number of rows in the corresponding measurement matrix $\Phi_f$, and also the number of weights of the adaptive filter in the proposed method is $[(500+80-1)/2]=289$, which is approximately half of the conventional method. The step size in both methods was $\mu=0.003$ and the initial values of the weights were set to zero.

Fig. 4 shows the mean of distortion versus different noise variances. We observe that the identification in the reduced dimension has lower distortion than the conventional method (approximately one octave in all noise variances). Moreover, as shown in Fig. 5, in the proposed method, the number of required iterations for convergence is approximately half of the conventional method. This attribute results in faster information acquisition. However, the total time for final identification in the proposed method also depends on the speed of the reconstruction algorithm. In contrast to the conventional method, in which the number of transmitted pilot samples is equal to the number of iterations, in the proposed method, it is $q$ times of the iteration numbers. Hence, in this experiment, the required pilots in both methods are approximately equal (Fig. 5).

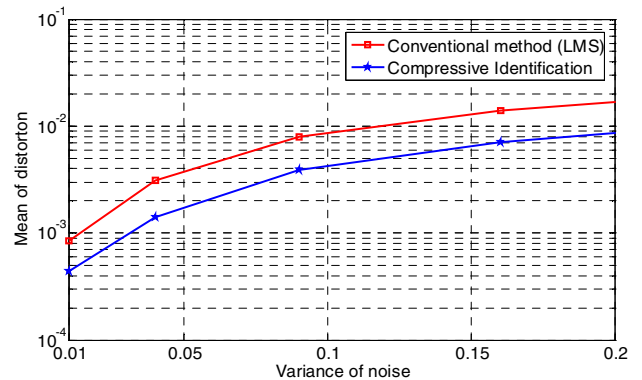

Figure 4. Mean of distortion in the time and compressed domain.

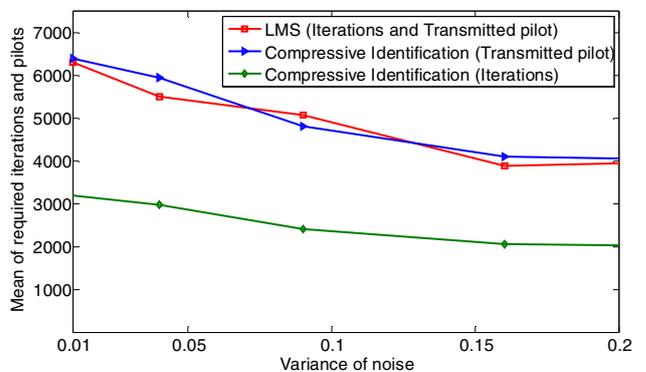

Figure 5. Number of iterations and transmitted pilots in the conventional and compressive identifications, $q=2$.

## V. DENOISING IN THE RECOVERY STAGE

In this experiment, we consider the potential performance of the proposed method in the final identification of the sparse systems (i.e. by considering the reconstruction stage). The first experiment was repeated for various levels of sparsity where noise variance was kept constant ($\sigma^2 = 0.01$). The parameter of ZA-LMS was tuned such that it yields the minimum possible distortion at $k=40$ and $\mu = 0.003$. We found $\rho = 2.5 \times 10^{-5}$ as the best value. Fig. 6 shows the convergence curve. Although, ZA-LMS has slightly less steady state distortion than the proposed compressive method, but our method converges with higher speed.

Fig. 7 compares the mean of steady state distortion of different implementations for various sparsity levels. Like the first experiment, the identification distortion of the reduced dimension (before reconstruction) is approximately half of the conventional method for all values of *k*. Although, in all depicted levels of sparsity, ZA-LMS has less distortion than the compressive identification, but the final identification in the proposed method which deploys the denoising performance of the sparse reconstruction algorithm, surpasses ZA-LMS in the low sparsity levels. In this experiment, we used the Bayesian algorithm with Laplacian priori at the recovery stage [12]. Of course, an algorithm which can exploit the structure of random filter measurement in the recovery process, would have better performance.

Fig. 8 compares the computational complexity of the different methods considering the required processing times in the given sparsity levels. Our simulations were performed in MATLAB 7.6 environment using a Dual-Core 2.7GHz processor with 2GB of RAM, under Microsoft Windows 7 operating system. Although our method consists of two stages, but it has less complexity than the ZA-LMS at the certain numbers of non-zero values.

## VI. CONCLUSION

We proposed a new efficient method for sparse system identification which introduces a new complexity-performance trade-off in this field. The proposed method is based on identifying the compressed version of the system that is achieved by interpolation and consequently filtering at the transmitter side and decimation at the receiver front end. We experimentally showed that an appropriate sparse reconstruction algorithm of compressed sensing theory can reduce the amount of identification distortion at the recovery stage. Certainly, an algorithm which is able to exploit the structure of random filter in the recovery process would have better denoising performance in the proposed strategy. In addition, exploiting any structure of desired sparse system will be easily possible at the reconstruction stage of the proposed approach.

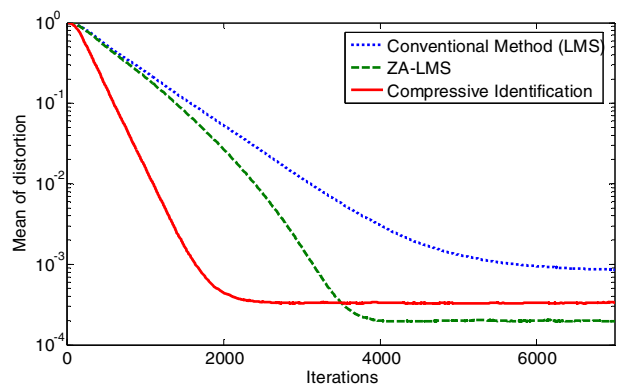

Figure 6. Convergence curve for three different methods, *N*=500, *k*=40.

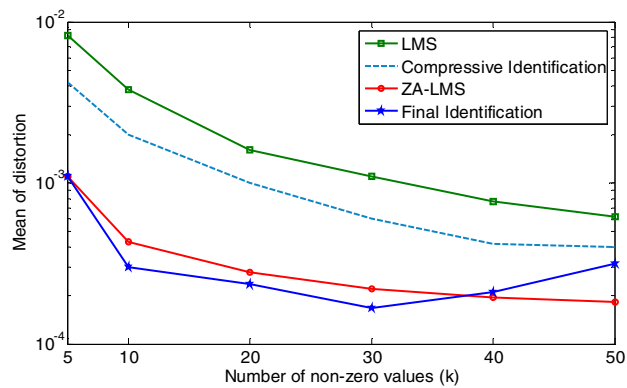

Figure 7. Potentially denoising performance at the proposed method.

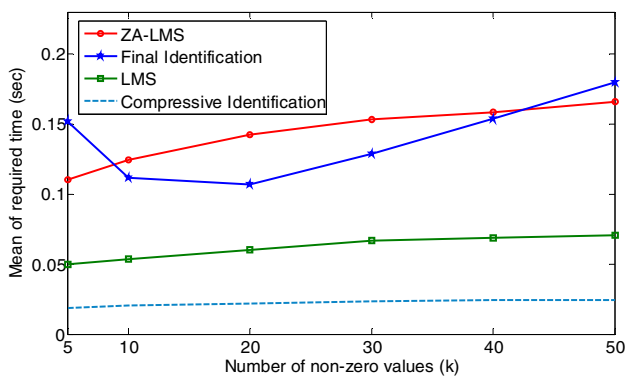

Figure 8. Processing times of the various methods.